\documentclass[12pt]{iopart}

\pdfoutput=1
 
\usepackage{latexsym}
\usepackage{graphics}
\usepackage{graphicx}
\usepackage{amssymb}
\usepackage{bm}
 
\begin{document}

\title{Ultracold Atoms as a Target: Absolute Scattering Cross-Section Measurements}

\author{P. W{\"u}rtz, T. Gericke, A. Vogler, and H. Ott}

\address{Institut f{\"u}r Physik, Johannes Gutenberg-Universit{\"a}t, 55099 Mainz, Germany \\ Research Center OPTIMAS, Technische Universit{\"a}t Kaiserslautern, 67663 Kaiserslautern, Germany}

\ead{ott@physik.uni-kl.de}

\begin{abstract}
We report on a new experimental platform for the measurement of absolute scattering cross-sections. The target atoms are trapped in an optical dipole trap and are exposed to an incident particle beam. The exponential decay of the atom number directly yields the absolute total scattering cross-section. The technique can be applied to any atomic or molecular species that can be prepared in an optical dipole trap and provides a large variety of possible scattering scenarios.
\end{abstract}

\maketitle

\section{Introduction}
A gaseous atomic target with very low momentum spread is an ideal starting point for atomic scattering experiments. This was demonstrated with great success by the invention of the COLTRIMS (cold target recoil ion momentum spectroscopy) technique \cite{Doerner2000}. While in COLTRIMS experiments, the target is an atomic beam with low transverse momentum spread, the advent of laser cooling and trapping has provided a related platform. It is known as MOTRIMS (magneto-optical trap recoil ion momentum spectroscopy) \cite{VanderPoel2001,Flechard2001,Turkstra2001}, and uses an atomic cloud as target which is cooled in all three spatial dimensions with a magneto-optical trap. The achievable temperature of $T\approx$ 100$\,\mu$K corresponds to an energy spread of only 10\,neV. 

The above mentioned experiments focus on charged reaction products which can be detected with a position sensitive micro-channel plate. The inclusion of scattering processes involving neutral reaction products is possible if one looks, e.g., at the temporal evolution of the target. This approach has the benefit that absolute cross-sections can be measured. In this context, the atom loss of a MOT under electron bombardment has enabled the measurement of the total scattering cross-section and the total ionization cross-section for electrons on rubidium atoms at electron energies up to 500\,eV \cite{Schappe1995,Schappe1996}.

In this work, we discuss the extension of this principle to a target of ultracold atoms which are held in an optical dipole trap. We give a first example of this technique measuring the total electron-rubidium scattering cross-section at energies between 1.7\,keV and 6\,keV. We assess the new possibilities of this experimental platform and the additional benefits compared to the preparation of the atoms in a MOT. 

\begin{figure}[htbp]
\centering
\includegraphics[width=7cm]{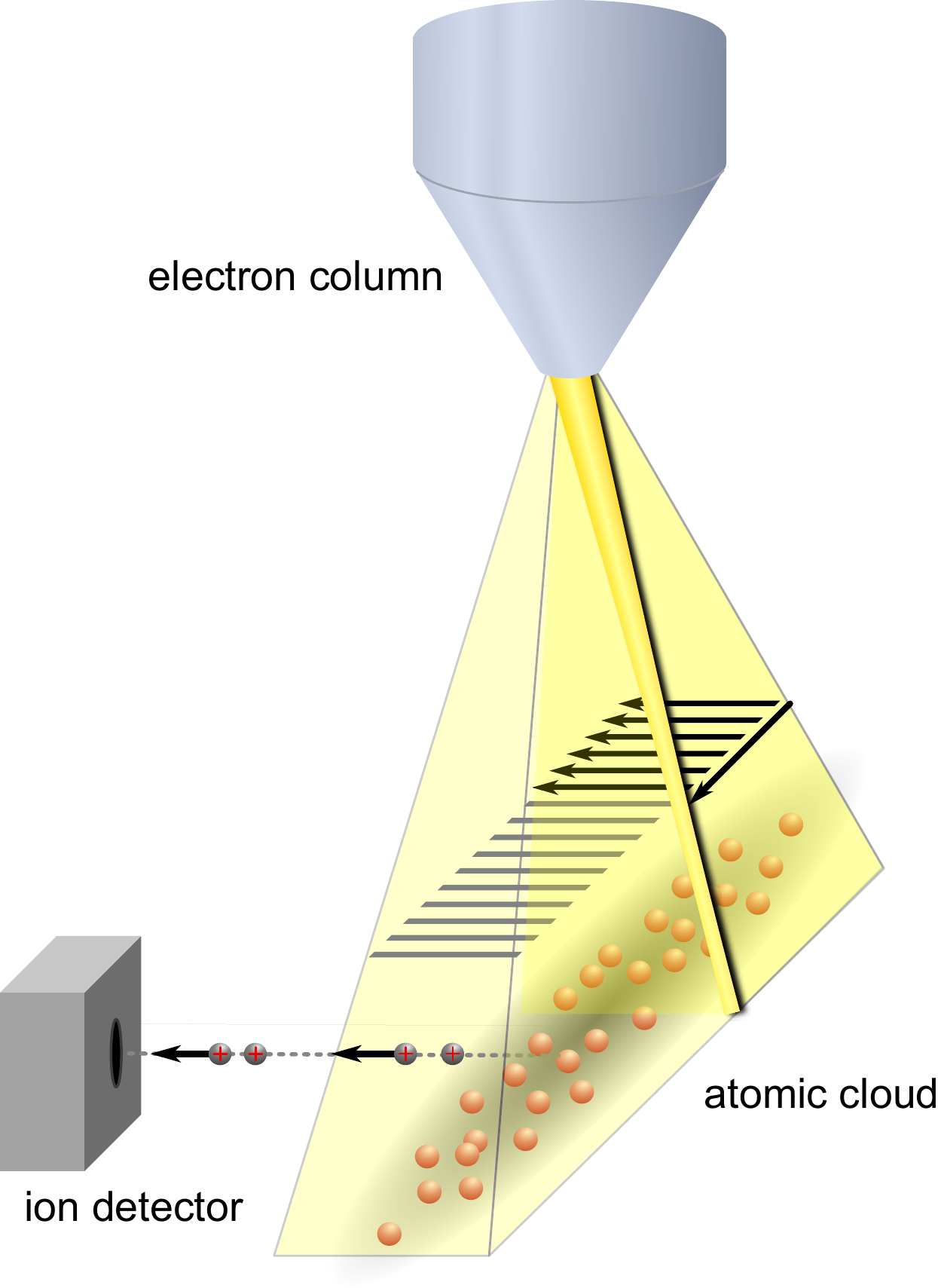}
\caption{Working principle of the cross-section measurement. A focussed electron beam repeatedly scans a rectangular area. The target atoms are trapped in an optical dipole trap and exposed to the electron beam. During the exposure, the number of target atoms decays. From the decay constant the scattering cross-section is infered.}
\label{fig:working_principle}
\end{figure}

\section{Measurement principle}

The measurement of absolute scattering cross-sections is of great importance for a quantitative comparison between experiment and theory. There are two different experimental strategies for their determination. In the first approach, the absolute density of target atoms has to be known. Then, it is sufficient to measure the relative number of scattered projectiles. The second strategy is reversed and requires the knowledge of the flux density of the incident projectiles. Then, the relative decay of the number of target atoms is sufficient to extract the total cross-section. This strategy can be used in crossed beam experiments or in experiments involving a gaseous target which is fixed in space. In both strategies, the spatial overlap integral between the projectiles and the target has to be determined as well. This task is simplified if the incoming flux density $\Phi$ of projectiles is spatially homogeneous and if the target - which we assume to be fixed in space - is completely immersed in the incoming projectiles. Then, the number of target atoms $N$ evolves according to

\begin{equation}
\dot{N}=-\Phi \sigma_{\mathrm{tot}}N-\gamma_{\mathrm{bg}}N=-\gamma N - \gamma_{\mathrm{bg}}N
\label{eq:1}
\end{equation}

Here, $\sigma_{\mathrm{tot}}$ is the total scattering cross-section and $\gamma_{\mathrm{bg}}$ accounts for an additional loss channel which might be present in the experiment. The value of $\gamma_{\mathrm{bg}}$ must be measured separately. Eq.\,\ref{eq:1} is valid when each scattering process leads to the loss of exactly one target atom. This requires that the trap which holds the target atoms is shallow enough to let every scattered target atom escape. Furthermore, collisions in which a scattered target atom removes another atom on its way out of the target have to be negligible. The solution of Eq.\,\ref{eq:1} is an exponential decay of the number of target atoms. The total scattering cross-section $\sigma_{\mathrm{tot}}$ is directly given by the decay constant $\gamma$ devided by the flux density $\Phi$. This principle has been experimentally demonstrated with rubidium atoms in a MOT which were exposed to an electron beam with energies up to 500\,eV\,\cite{Schappe1995, Schappe1996}. In an analogous approach, a light beam which intersects a cloud of trapped negative ions has recently been used to measure absolute photodetachment cross-sections \cite{Hlavenka2009}.

\section{Experimental procedure}

In our experiment, we extend this approach to an ultracold gaseous target which is prepared in an optical dipole trap. Starting from MOT, we load $2\times10^6$ rubidium atoms in an optical dipole trap. The dipole trap is formed by a focussed CO$_2$ laser beam with a waist of 30\,$\mu$m. After an additional stage of forced evaporation we obtain samples of $1-3\times10^5$ rubidium atoms at a temperature between 50\,nK and 200\,nK. Below 150\,nK the atoms form a Bose-Einstein condensate. This temperature range corresponds to a trap depth between 30 an 140 peV. The details of the experimental setup can be found in \cite{Gericke2008,Wuertz2009}. 

\begin{figure}[htbp]
\centering
\includegraphics{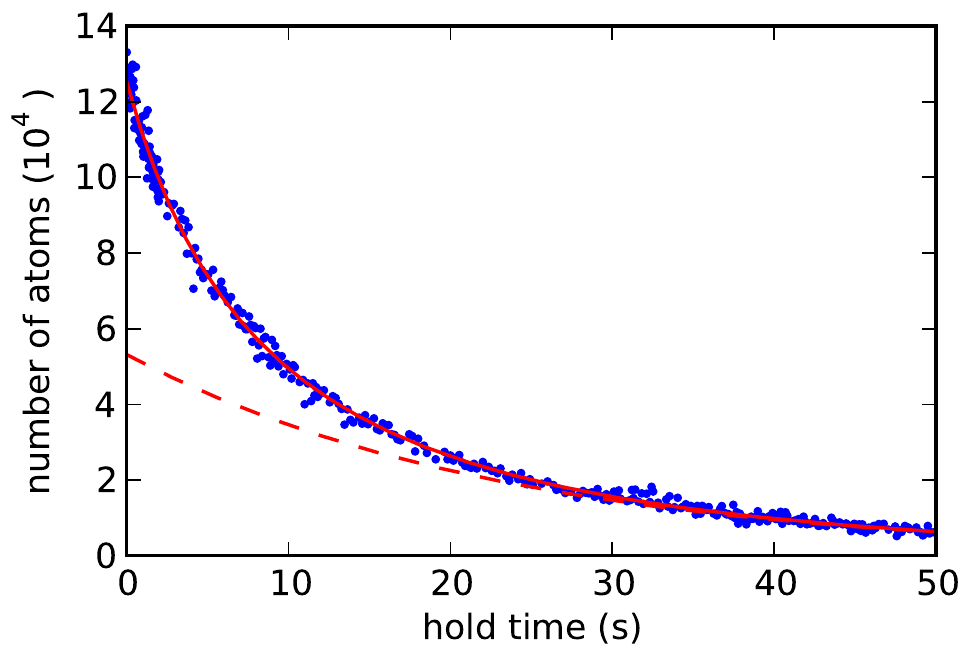}
\caption{Mesurement of the trap lifetime. Without exposure to the electron beam, the number of atoms decays within several 10 seconds. After a slightly faster initial decay which is due to evaporative cooling and two-body losses, we find an exponential decay with a time constant of 25\,s. This decay adds to the exponential decay upon exposure (see Eq.\,\ref{eq:1}). As the time scale of this decay is very long, the resulting correction to the determined scattering cross-section is small.}
\label{fig:trap_lifetime}
\end{figure}

The collisional system is completed by an incident electron beam originating from an electron column. As the experimental setup has been developed in the context of scanning electron microscopy of ultracold quantum gases \cite{Gericke2008}, the electron beam can be focussed down to about 100\,nm diameter and has an energy between 1.7\,keV and 6\,keV. Typical beam currents vary between 10\,nA and 1\,$\mu$A, depending on energy and beam diameter. The cloud of target atoms is cigar shaped with a radial extension of 10\,$\mu$m and an axial extension of 100\,$\mu$m. After the preparation stage we switch on the focussed electron beam and repeatedly scan an area $A$ which is about three times as large as the size of the cloud. Each one of these frames takes 18\,ms and consists of 400 parallel lines which are oriented perpendicular to the long axis of the cloud (see Fig.\,\ref{fig:working_principle}). The scanning speed within each line and the propagation speed of the lines along the axial direction of the cloud is much faster than the motion of the atoms. Therefore, the electron beam crosses an unperturbed cloud during one frame and the action of the electron beam can be integrated over the frame time. We make one hundred consecutive frames, resulting in a total exposure time of 1.8\,s. At the end of the exposure the cloud is depleted almost entirely. The total experimental cycle has a duration of 15\,s.

When the diameter of the electron beam is much larger than the distance between two neighboring scan lines it is obvious that the integration of the current density over the frame time results in an effectively homogeneous current density. However, for a tightly focussed electron beam where the electron beam diameter is smaller than the distance between two neighboring scan lines, the current density after integration is strongly inhomogeneous. Nevertheless, it can be considered homogeneous provided that (i) the target density is sufficiently constant over the distance between two neighboring lines and (ii) the dwell time at a certain position is short enough that only a small fraction of the target atoms is lost, i.e., the number of scattered target atoms is linear in dwell time. Both conditions are fulfilled for our experimental parameters . We measure the area over which the electron beam is scanned with help of a two-dimensional optical lattice which sets a regular structure with 600\,nm period. Imaging the atoms in the lattice \cite{Wuertz2009} allows us to calibrate the scan system of the electron column. A Faraday cup measures the total beam current $I$ and we get the incoming flux density as

\begin{equation}
\Phi=\frac{I}{eA}\,,
\label{eq:2}
\end{equation}

where $e$ is the electron charge. During the exposure, electron impact ionization leads to a continuous production of ions which we detect with a channeltron. The number of produced ions is recorded and binned for each frame. As the motion of the electron beam is much faster than the atomic motion the binned signal is proportional to the total atom number at the beginning of each frame. We repeat the experiment cycle several hundred times and sum the signal over all runs. Collisions with the background gas limit the lifetime of the target atoms in the optical dipole trap and constitute an additional decay as introduced in Eq.\,\ref{eq:1}. We measure the corresponding decay constant $\gamma_{\mathrm{bg}}$ in a separate measurement (see Fig.\,\ref{fig:trap_lifetime}).

\begin{figure}[htbp]
\centering
\includegraphics{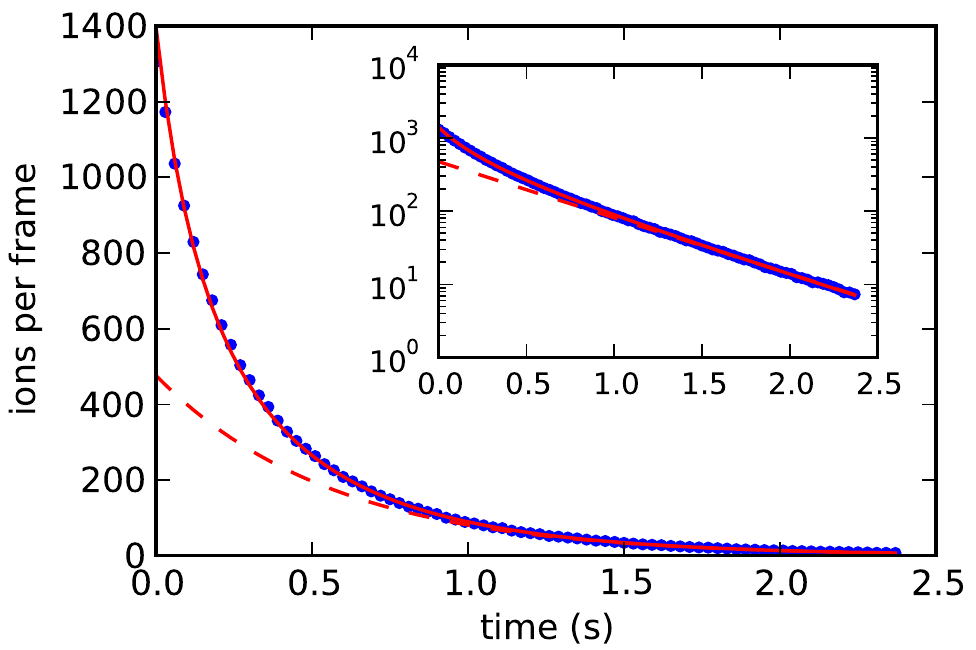}
\caption{Decay of the atom number during the illumination with the electron beam. The number of detected ions per frame is plotted versus time. The red dashed line is a fit with a pure exponential decay. The red solid line is a fit based on Eq.\,\ref{eq:3}. Inset: The logarithmic plot reveals the pure exponential decay which sets in after 1\,s.}
\label{fig:decay_curve}
\end{figure}

\section{Results and discussion}

A typical decay curve of the atom number is presented in Fig.\,\ref{fig:decay_curve}. The data was taken at a beam energy of 6\,keV which corresponds to the standard working point of the electron column. According to Eq.\,\ref{eq:1}, the decay of the atom number should be exponential. We find that the exponential decay sets not in until a substantial fraction of the atoms is already lost. The deviation from Eq.\,\ref{eq:1} is due to secondary processes, where a scattered target atom or a produced ion can remove another atom from the target. These processes can be modelled with an additional decay term which is quadratic in the atom number

\begin{equation}
\dot{N}=-\gamma N-\gamma_{\mathrm{bg}}N-\beta N^2.
\label{eq:3}
\end{equation}

\begin{figure}[htbp]
\centering
\includegraphics[width=12cm]{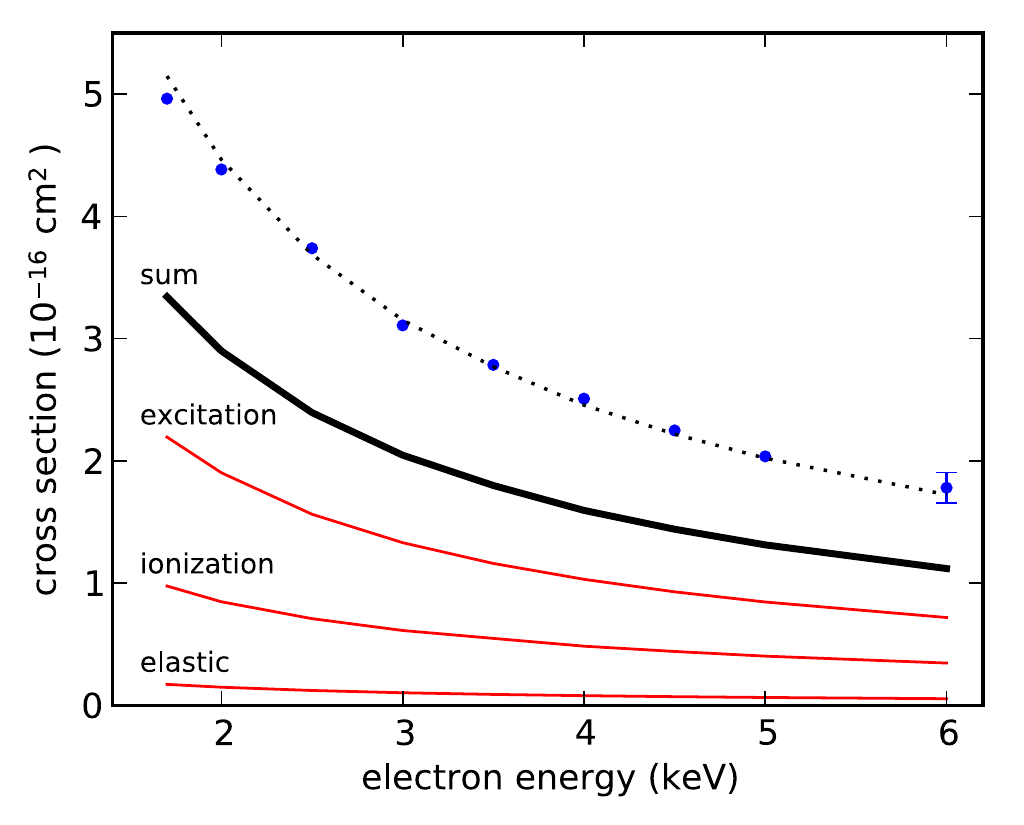}
\caption{Absolute total scattering cross-section measurement. The total scattering cross-section for electrons on rubidium atoms in an energy range between 1.7\,keV and 6\,keV has been measured (blue points). The solid red lines show the theoretical predictions for elastic scattering, electron impact ionization and electron impact excitation. The sum of all three contributions is shown as black line (the dotted line is the sum multiplied with a factor of 1.52, see text). The cross section at 6\,keV energy has been measured several times for different experimental parameters. The plotted error bar shows the spread of these measurements.}
\label{fig:cross_sections}
\end{figure}

The coefficient $\beta$ describes the strength of these processes. As can be seen from Fig.\,\ref{fig:decay_curve} the agreement with the data is very good over the whole exposure time. This confirms the presence of secondary processes. We attribute them to cold ion-atom collisions, as only this collisonal system has a sufficient cross-section to explain the frequency of these processes \cite{Cote2000}. From the fit to Eq.\,\ref{eq:3} we extract the decay constant $\gamma$, and together with the previously measured flux density $\Phi$ we deduce the absolute total scattering cross-section $\sigma_{\mathrm{tot}}$. We have performed the measurement for incident energies between 1.7\,keV and 6\,keV. The results are summarized in Fig.\,\ref{fig:cross_sections}. The uncertainty of our measurement has been evaluated by averaging over several datasets taken at an energy of 6\,keV. We have varied the beam diameter, the size of the scanned area and have changed between condensed and thermal samples. At 6\,keV energy, we determine a cross section of $\sigma_{\mathrm{tot}}=(1.78 \pm 0.14) \times 10^{-16}$\,cm$^2$, corresponding to an uncertainty of $8\,\%$.

\begin{figure}[htbp]
\centering
\includegraphics{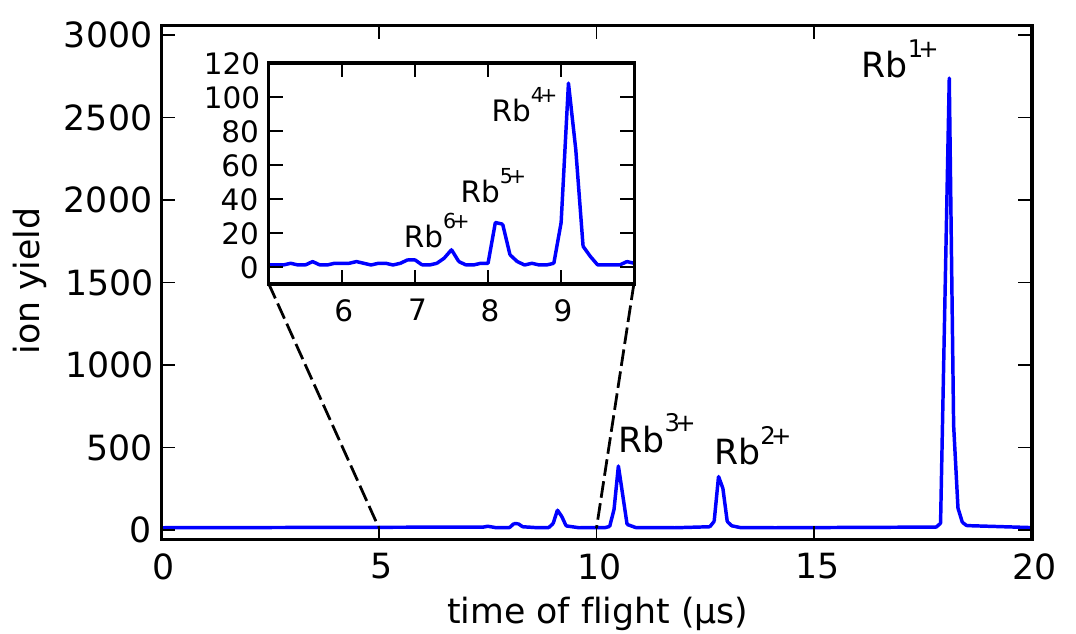}
\caption{Time of flight spectrum of the produced ions. 80\,\% of the produced ions are singly charged. We also find higher charge states up to Rb$^{7+}$.}
\label{fig:tof_spectrum}
\end{figure}

While there are no experimental data available in this energy range, we can compare our results to theoretical predictions. The total scattering cross-section has three contributions: 

\begin{itemize}

\item{{\bf Electron impact excitation:}. The general expression for the differential cross-section in first Born-approximation is given by \cite{Inokuti71}

\begin{equation}
\frac{d \sigma_n}{d \Omega} = c({\bf k},{\bf k'},Z) \left| \langle  n |\sum_{j=1}^Z  e^{i{\bf q}{\bf r}_j} |0\rangle\right|^2.
\end{equation}

Here, $c({\bf k},{\bf k'},Z)$ is a prefactor that depends on the wave vectors ${\bf k}$ and ${\bf k'}$ of the incoming and outgoing electron and the total number of target electrons $Z$, whose positions are denoted by ${\bf r}_j$. The momentum transfer is defined as ${\bf q} = {\bf k} - {\bf k'}$  and $|0\rangle$ and $|n\rangle$ are the initial and final state of the target. For vanishing momentum transfer (${\bf q}\rightarrow 0$), the matrix element approaches that for optical transitions $\langle  n |  i {\bf q}{\bf r}  |0\rangle$. These kind of collisions constitute the dominant excitation channel and are referred to as ''dipole regime''. For rubidium, excitation on the 5s - 5p resonance line is the most important contribution. In Ref.\,\cite{Chen1978} an empirical formula for the cross-section of the 5s - 5p transition has been given, based on the first Born-approximation. We estimate that the excitation to higher lying states (5s - np, n=6,7,...) cumulate to about 10\,\% of the cross-section of the strong resonance line.}

\item{{\bf Elastic scattering:} We employ an empirical formula, which is supposed to be applicable to all elements and all energies above 100\,eV \cite{Browning1994}. Within our energy range, the elastic cross-section is about 10\,\% of the impact excitation cross-section. In all elastic or exciting collisions, the momentum tranfer to the atom is substantially larger than the trap depth, and every scattering process leads to the loss of the atom.}

\item{{\bf Electron impact ionization:} In the context of scanning electron microscopy of ultracold quantum gases, electron impact ionization is the relevant scattering mechanism as it produces a detectable signal. The ionization process includes singly and multiply charged ions. A time of flight spectrum, presented in Fig.\,\ref{fig:tof_spectrum}, shows the relative weights of these contributions: more than 80\,\% of the ions are singly charged. While experimental data for the total ionization cross-section is availible only up to an energy of 500\,eV \cite{Schappe1996}, theoretical calculations in plane-wave Born-approximation have been made up to an energy of 10\,keV\,\cite{Bartlett2004}.} 

\end{itemize}

For a quantitative comparison we plot the three contributions together with their sum in Fig.\,\ref{fig:cross_sections}. At all data points the theoretical prediction differs form the experimental result by an almost constant factor ($1.52\pm$0.04). This suggests that either the experimental data systematically overestimates the cross-section or the theoretical predictions miss a substantial amount of scattering processes. 

The measurement could in principle be influenced by the presence of the optical dipole trap. We have tested this in the following way: we direct the electron beam into the center of the cloud and record the ion signal for 100\,$\mu$s. We then repeat the experiment switching off the dipole trap during this time. Within the first 100\,$\mu$s, the expansion of the cloud can be neglected and the electron beam interacts with a cloud of the same density distribution but without the presence of any trapping light. We find the same ion signal and can therefore exclude such an effect. Varying the parameters of the electron beam and the size and properties of the scan area, we can also exclude any influence stemming from the specific realization of the experiment. We can furthermore exclude an inaccurate determination of the size of the scan area, as the two-dimensional optical lattice provides a perfectly periodic ruler. A potential source for a systematic error could be the determination of the electron beam current. We use a Faraday cup which can be biased and which has an internal transverse magnetic field in order to prevent elastically backscattered electrons from escaping from the cup. We conclude that a systematic error originating from the Faraday cup is unlikely, however we have currently no means to independently calibrate it as it is an integral part of the setup.

The above presented theoretical predictions are based on simplifying assumptions such as the first Born-approximation. This might lead to a systematic underestimation of the cross-sections. In addition, more complicated excitation channels, such as the excitation of inner shell electrons, the inclusion of optically forbidden transitions or the excitation of more than one electron, have not been accounted for in our simple model. Our results might indicate that these channels also contribute significantly to the toal cross-section. Apart from the normalization factor, the data trend shows very good agreement between experiment and theory. The good quality of the measurement is further confirmed by the small uncertainty. We therefore conclude that the presented approach is suitable for high precision measurements of absolute scattering cross-sections.

Compared to previously reported experiments using a magneto-optical trap \cite{Schappe1995,Schappe1996} several differences are apparent. In our approach, no switching of magnetic fields is necessary, as the dipole trap is extremely shallow. The experiment can be performed continously until the target is fully depleted. Thus, a single experimental run is already sufficient to derive the cross-section. As only the relative atom number is important, individual experimental runs can be summed without normalization. Finally, the atoms in the dipole trap can be polarized, which allows for spin-resolved scattering experiments.  

\section{Outlook}

We have described a new experimental platform for the measurement of absolute scattering cross-sections based on optically trapped atoms. We have demonstrated the principle studying electron rubidium collisions at high incident energies. Even though the measurement principle relies on the use of optical trapping fields, a surprisingly large variety of scattering scenarios is feasible. Based on the actual status of cold atom physics, including the recent developments in non-optical cooling techniques and molecule formation, we can identify a number of interesting scattering scenarios and applications for the future:  

\begin{itemize}

\item{{\bf Low energy electron-atom collisions:} The combination of laser cooling and subsequent photoionization is well suited to produce ions and electrons with extremely small initial energy spread \cite{Hanssen2008}. Implementing two neighboring dipole traps, one of which is used as an electron source and the other is used as a target, allows to study low energy electron-atom collisions with unprecedented energy resolution. Polarizing the atoms in both traps provides full control over the spins of the incoming electrons and the target atoms.}

\item{\bf Low energy ion-atom collisions:} The same holds for the investigation of ultracold ion-atom collisions. As the recoil energy of the ion is negligible in photoionization, an even higher energy resolution should be feasible.

\item{{\bf Electron-Rydberg atom collisions:} Recently, a new form of molecular binding mechanism has been identified for ultracold Rydberg atoms \cite{Bendkowsky2009}. Extending these studies by exciting atoms to Rydberg states and exposing them to a low energy electron beam can give more insight in these phenomena. It also complements various studies of plasma physics with Rydberg atoms \cite{Robinson2000}.}

\item{{\bf Molecular targets:} Feshbach resonances can be used to produce ground state molecular dimers from alkali atoms \cite{Ferlaino2009}. The control over all internal and external degress of freedom offers the unique opportunity to produce a well defined molecular target, where fundamental electron-molecule or ion-molecule collisions can be studied. More complex molecules could be available soon combining Stark deceleration and subsequent optical trapping.}

\end{itemize}

\ack
We gratefully acknowledge financial support from the DFG under Grant No. Ot 222/2-3.

\end{document}